\documentclass[iop]{emulateapj}
\usepackage{hyperref}

\shorttitle{Core Mass Function in the Massive Protocluster G286.21+0.17}
\shortauthors{Cheng et al.}

\begin{document}

\title{The Core Mass Function in the Massive Protocluster G286.21+0.17 revealed by ALMA}

\author{Yu Cheng$^1$, Jonathan C. Tan$^{1,2,3,4}$, Mengyao Liu$^{1,4}$, Shuo Kong$^{5}$, Wanggi Lim$^{1}$, Morten Andersen$^{6}$, Nicola Da Rio$^{4}$}
\affil{$^1$Dept. of Astronomy, University of Florida, Gainesville, Florida 32611, USA\\
$^2$Dept. of Physics, University of Florida, Gainesville, Florida 32611, USA\\
$^3$Dept. of Space, Earth \& Environment, Chalmers University of Technology, Gothenburg, Sweden\\
$^4$Dept. of Astronomy, University of Virginia, Charlottesville, Virginia 22904, USA\\
$^5$Dept. of Astronomy, Yale University, New Haven, CT 06511, USA\\
$^6$Gemini Observatory, Casilla 603, La Serena, Chile}


\begin{abstract}
We study the core mass function (CMF) of the massive protocluster
G286.21+0.17 with the Atacama Large Millimeter/submillimeter Array via
1.3~mm continuum emission at a resolution of 1.0\arcsec\ (2500~au). We
have mapped a field of 5.3\arcmin$\times$5.3\arcmin\ centered on the
protocluster clump. We measure the CMF in the central region,
exploring various core detection algorithms, which give source numbers
ranging from 60 to 125, depending on parameter selection. We estimate
completeness corrections due to imperfect flux recovery and core
identification via artificial core insertion experiments.  For masses
$M\gtrsim1\:M_\odot$, the fiducial dendrogram-identified CMF can be
fit with a power law of the form
${\rm{d}}N/{\rm{d}}{\rm{log}}M\propto{M}^{-\alpha}$ with
$\alpha\simeq1.24\pm0.17$, slightly shallower than, but still
consistent with, the index of the Salpeter stellar initial mass
function of 1.35. Clumpfind-identified CMFs are significantly
shallower with $\alpha\simeq0.64\pm0.13$. While raw CMFs show a peak
near $1\:M_\odot$, completeness-corrected CMFs are consistent with a
single power law extending down to $\sim 0.5\:M_\odot$, with only a
tentative indication of a shallowing of the slope around
$\sim1\:M_\odot$. We discuss the implications of these results for
star and star cluster formation theories.
\end{abstract}

\keywords{stars: formation -- ISM: clouds}

\section{Introduction}\label{S:intro}

The stellar initial mass function (IMF) is of fundamental importance
throughout astrophysics. However, in spite of much progress in
measuring the IMF (see reviews of, e.g., Bastian et al. 2010; Kroupa
et al. 2013), its origin and environmental dependence are still under
active debate. Stars are known to form from cold dense cores in
molecular clouds. These ``prestellar cores'' can be defined
theoretically as gravitationally-bound, local density maxima that
collapse via a single rotationally-supported disk into a single star
or small $N$ multiple. In the context of Core Accretion models (e.g.,
Padoan \& Nordlund 2002, 2007; McKee \& Tan 2003; Hennebelle \&
Chabrier 2008; Kunz \& Mouschovias 2009), the stellar mass is assumed
to be related to the mass of its parental core, modulo a relatively
constant core to star formation efficiency, $\epsilon_{\rm core}$,
perhaps set mostly by outflow feedback (Matzner \& McKee 2000; Zhang
et al. 2014), with radiative feedback expected to influence only the
most massive stars (Tanaka et al. 2017). In this framework, we expect
the IMF to be strongly influenced by the prestellar CMF, i.e., the
PSCMF. However, there are alternative models, especially Competitive
Accretion (Bonnell et al. 2001; Bate 2012), which explain the IMF
without a CMF that extends to higher masses. Therefore, the study of
the CMF, and ideally the PSCMF, is crucial for understanding the
origin of the IMF and its connection to the large-scale physical and
chemical conditions of molecular clouds.

Early observations based on submillimeter dust continuum emission
(e.g., Motte et al. 1998; Testi \& Sargent 1998; Johnstone et
al. 2000) found evidence for an
approximately log-normal CMF peaking near $\sim1\:M_\odot$, with a
power law tail at higher masses of the form
\begin{equation}  
\frac{{\rm d} N}{{\rm d} {\rm log} M}\propto{M}^{-\alpha}.
\label{eq:pl}
\end{equation}
These studies found values of $\alpha\simeq1.0$ to 1.5, based on
samples of several tens of sources. In this form, the Salpeter (1955)
$\gtrsim1\:M_\odot$ power law fit to stellar masses has an index
$\alpha=1.35$, indicating a potential similarity of the CMF and
IMF. Alves et al. (2007) used near-infrared dust extinction to
characterize about 160~cores to find similar results, with the peak of
the CMF now better measured close to $1\:M_\odot$ and the CMF reported
to be a simple translation of the IMF requiring $\epsilon_{\rm
  core}\simeq0.3$.
More recent results from the Gould Belt Survey with {\it{Herschel}},
{\it{Spitzer}} and JCMT have also detected samples of hundreds of
cores (e.g., Andr\'e et al. 2010; Sadavoy et al. 2010; Salji et
al. 2015; Marsh et al. 2016) and have added to the evidence for a
similarity in shape of the CMF and IMF.


Extending to more distant ($\gtrsim2\:$kpc), high-mass star-forming
regions has been more challenging, in particular requiring higher
angular resolution interferometric observations.
Beuther \& Schilke (2004; see also Rodon et al. 2012) reported a CMF
of 1.3~mm emission cores in IRAS~19410+2336 ($d\sim2\:$kpc) with
$\alpha\simeq1.5\pm0.3$, based on a sample of 24 sources ranging in
mass from $\sim2-25\:M_\odot$.  Bontemps et al. (2010) detected a
similar number of sources in Cygnus X ($d=1.7\:$kpc), but these were
identified from the follow-up of five quite widely-separated clumps,
so that the CMF was not derived from uniform mapping of a contiguous
region. Zhang et al. (2015) studied the core population via 1.3~mm
emission in the Infrared Dark Cloud (IRDC) G28.34 P1 clump
($d\simeq5\:$kpc) with ALMA, finding 38 cores. They concluded there
was a dearth of lower-mass ($\sim1-2\:M_\odot$) cores compared to the
prediction resulting from a scaling down to these masses with a
Salpeter mass function. Ohashi et al. (2016) studied the IRDC
G14.225-0.506 ($d=2\:$kpc) CMF via 3~mm emission with ALMA at
$\sim3\arcsec$ resolution, identifying 48 sources with the clumpfind
algorithm (Williams et al. 1994) from two separate fields. They
derived $\alpha=1.6\pm0.7$, with the masses ranging from
$1.5-22\:M_\odot$.


G286.21+0.17 (hereafter G286) is a massive protocluster associated
with the $\eta$ Car giant molecular cloud at a distance of
$2.5\pm0.3\:$kpc, in the Carina spiral arm (e.g., Barnes et al. 2010,
hereafter B10; Andersen et al. 2017). G286 has been claimed to be
$\sim 10^4\:M_\odot$ (B10), which would make it the most massive and
densest of the 300 HCO$^+$(1-0) clumps studied by Barnes et al. (2011)
and Ma et al. (2013), but an assessment of its dust mass from
{\it{Herschel}} imaging data suggests a lower mass of
$\sim2000\:M_\odot$ (Ma et al., in prep.). From modeling of HCO$^+$ and
H$^{13}$CO$^{+}$ spectra, B10 found a global infall rate
$\sim3\times10^{-2}\:M_\odot\:{\rm{yr}}^{-1}$, one of the largest such
infall rates yet measured.

Here we present the ALMA Band 6 (230~GHz) continuum observation of
G286 and an analysis of the CMF in this region. This paper is
organized as follows: in \S2 we describe the observational setup and
analysis methods; in \S3 we present our results, including an
exploration of different analysis techniques for identifying cores and
the resulting CMFs; in \S4 we discuss and summarize our conclusions.

\section{Observations and Analysis Methods}\label{S:obs}


\subsection{Observational Set-Up}

\begin{figure*}[ht!]
\epsscale{1.2}\plotone{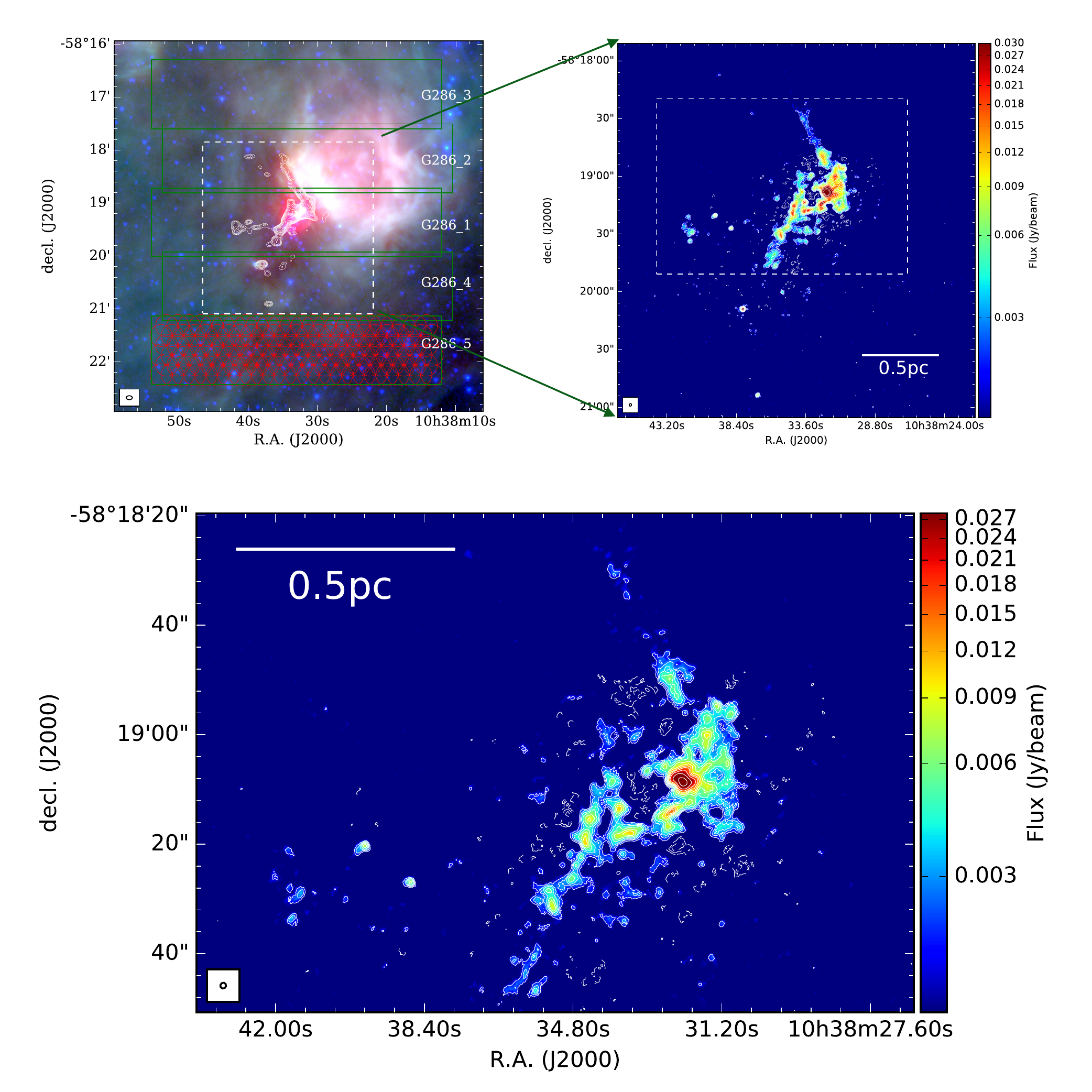}
\caption{
{\it (a) Top Left:} Three color image of G286 constructed by combining
{\it Spitzer} IRAC 3.6~$\mu$m (blue), 8.0~$\mu$m (green), and {\it
  Herschel} PACS 70~$\mu$m (red). White contours show ALMA 7-m array
image starting from 4$\sigma$. The G286 field is divided into five
strips, as shown by the green rectangles. Each strip is covered with
147 pointings of the 12-m array, illustrated for strip G286\_5 as an
example with red circles marking the FWHM field of view of each
pointing. The white dashed rectangle is the region shown in (b). {\it
  (b) Top Right:} Image with combined 12-m array and 7-m array data. The
resolution is 1.62\arcsec$\times$1.41\arcsec. The contour levels are
at (4, 8, 10, 12, 15, 20, 25, 30, 40, 50, 75, 100,
150)$\times0.45\:{\rm mJy}\:{\rm beam}^{-1}$ (color scale in
Jy~beam$^{-1}$). The white dashed rectangle is the region shown in
(c). {\it (c) Bottom:} Image with combined 12-m array and 7-m array
data, but now imaged at 1.07\arcsec$\times$1.02\arcsec. Our CMF
analysis is carried out for this region.}\label{fig:1}
\end{figure*}

The observations were conducted with ALMA during Cycle 3 (Project ID
2015.1.00357.S, PI: J. C. Tan), during a period from Dec. 2015 to
Sept. 2016. To map the entire field of G286
($\sim$5.3\arcmin$\times$5.3\arcmin), we divided the region into five
strips, denoted as G286\_1, G286\_2, G286\_3, G286\_4, and G286\_5,
each about $1\arcmin\:$ wide and $5.3\arcmin\:$ long and containing
147 pointings of the 12-m array. Figure~\ref{fig:1}a illustrates the
spatial extent of the five strips, together with red circles showing
the 12-m array mosaic footprints overlaid on strip G286\_5 as an
example. The position of field center is R.A.=10:38:33,
decl.=-58:19:22. We employed the compact configuration C36-1 to
recover scales between 1.5\arcsec\ and 11.0\arcsec. Additionally, a
35-pointing mosaic was performed for each strip using the 7-m array,
probing scales up to 18.6\arcsec. Total power observations of the
region were also carried out (relevant only for the line
observations).

Two scheduling blocks happened to be observed when the array
configuration was in a transition phase, i.e., moving from a very
extended configuration (C37/C38-1) to our proposed compact
configuration. Thus we obtained extra $uv$ coverage for two strips,
G286\_1 and G286\_2, where $\sim$90\% of the continuum emission is
located. This enables us to detect and characterize structures at a
higher resolution ($\sim$1\arcsec, 2500~au) in these regions, which
will be the focus of the results presented in this paper.

During the observations, we set the central frequency of the
correlator sidebands to be the rest frequency of the
$\rm{N_2D}^+$(3-2) line at $231.32\:$GHz for SPW0, and the
$\rm{C^{18}O}$(2-1) line at $219.56\:$GHz for SPW2, with a velocity
resolution of 0.046 and 0.048$\:\rm{km\:s}^{-1}$, respectively. The
second baseband SPW1 was set to $231.00\:$GHz, i.e., 1.30~mm, to
observe continuum with a total bandwidth of $2.0\:$GHz. The frequency
coverage for SPW3 ranges from 215.85 to $217.54\:$GHz to observe
DCN(3-2), DCO$^+$(3-2), SiO($v=0$)(5-4) and
$\rm{CH_3OH}(5_{1,4}-4_{2,2})$. The molecular line data from this
observation will be presented and analyzed in a future paper, while
here we focus on the results of the broad continuum band, i.e.,
tracing dust emission.

Both the 7-m and 12-m array data were calibrated with the data
reduction pipeline using {\it{Casa}} 4.7.0. The continuum visibility
data was constructed with all line-free channels. We performed imaging
with {\it tclean} task in {\it Casa} and during cleaning we combined
data for all five strips to generate a final mosaic map. The 7-m array
data was imaged using a Briggs weighting scheme with a robust
parameter of 0.5, which yields a resolution of
$7.32\arcsec\times4.42\arcsec\:$. For the combined data, we used the
same Briggs parameter. In addition, since we have extra $uv$ coverage
for part of the data, we also apply a 0.6\arcsec {\it uvtaper} to
suppress longer baselines, which results in
$1.62\arcsec\times1.41\arcsec\:$ resolution.

The lowest noise level in the image
varies from $0.2\:{\rm{mJy}}\:{\rm{beam}}^{-1}$ to
$0.46\:{\rm{mJy}}\:{\rm{beam}}^{-1}$, depending on which strip is
being considered. The $1\sigma$ noise of the central strip is
$0.45\:{\rm{mJy}}\:{\rm{beam}}^{-1}$. We also do the cleaning
separately for the central two strips with a smaller {\it uvtaper}
value to utilize the long baseline data, which results in a resolution
of 1.07\arcsec$\times$1.02\arcsec. The 1$\sigma$ noise level in this
image is $0.45\:{\rm{mJy}}\:{\rm{beam}}^{-1}$.

\subsection{Core Identification}

To study the CMF we first need to identify the ``cores.'' 
A variety of algorithms have been used to detect and characterise
dense cores in previous studies of continuum maps (e.g.,
Williams et al. 1994; Kramer et al. 1998; Rosolowsky et al. 2008), and
in practice, the results in terms of core number and statistical
properties can vary with the different algorithms and input parameters
(e.g., Pineda et al. 2009).
To understand how the derived CMF depends on these identification
methods, we thus adopt two well-documented and widely used algorithms
to analyse our data and test the effects of variation of their
parameters.

\subsubsection{The Dendrogram Method}

The dendrogram algorithm is described by Rosolowsky et al. (2008)
and implemented in {\it astrodendro}. The dendrogram is an abstraction
of the changing topology of the isosurfaces as a function of contour
level. This method can describe hierarchical structures in a 2-D or
3-D datacube. There are two types of structures returned in the
results: leaves, which have no sub-structure; and branches, which can
split into multiple branches or leaves. Here we only use the leaf
structure as a representation of dense cores.

There are three main parameters in this algorithm: $F_{\rm{min}}$,
$\delta$, and $S_{\rm{min}}$. First, $F_{\rm{min}}$ is the minimum
value to be considered in the dataset. In the fiducial case we adopt
$F_{\rm{min}}=4\sigma$. Second, $\delta$ describes how significant a
leaf has to be in order to be considered as an independent entity. We
adopt a fiducial value of $\delta=1\sigma$, which means a core must
have a peak flux reaching $5\sigma$ above the noise. The minimum area
a structure must have to be considered as a core is given by
$S_{\rm{min}}$. In general the size of the beam is a good choice, but
in a crowded field a detected core can be smaller than one beam size
due to blending, especially when a large value of $F_{\rm min}$ is
used. We thus set $S_{\rm{min}}=0.5S_{\rm beam}$ as our fiducial
choice. We will also explore the effects of varying these choices of
$F_{\rm min}$, $\delta$, and $S_{\rm min}$.

\subsubsection{The Clumpfind Method}

The clumpfind algorithm (Williams et al. 1994) works by first
contouring the data at a multiple of the rms noise of the observation,
then searching for peaks of emission that locate the structure, then
following them down to lower intensities. It was designed to study
molecular clouds using 3-D datacubes and has also been widely used to
describe dense cores (e.g., Reid \& Wilson 2005; Pineda et al. 2009).

The most sensitive parameters for clumpfind are the lowest contour
level ($F_{\rm min}$) and level spacing ($\Delta$). $F_{\rm min}$ is
the same as that in the dendrogram method, and we adopt $4\sigma$ as a
fiducial value. $\Delta$ refers to the contour level spacing and hence
is somewhat different from the $\delta$ parameter of the dendrogram
method. We choose $\Delta=3\sigma$ in the fiducial case, similar to
previous implementations in the literature. 
As with the dendrogram method, cores are requires to have a minimum
area $S_{\rm min}$, and we adopt $S_{\rm{min}}=0.5S_{\rm beam}$ as a
fiducial threshold. Again, we investigate the effects of variations in
the values of $F_{\rm min}$, $\Delta$, and $S_{\rm min}$.

\begin{figure*}[ht!]
\epsscale{1.2}\plotone{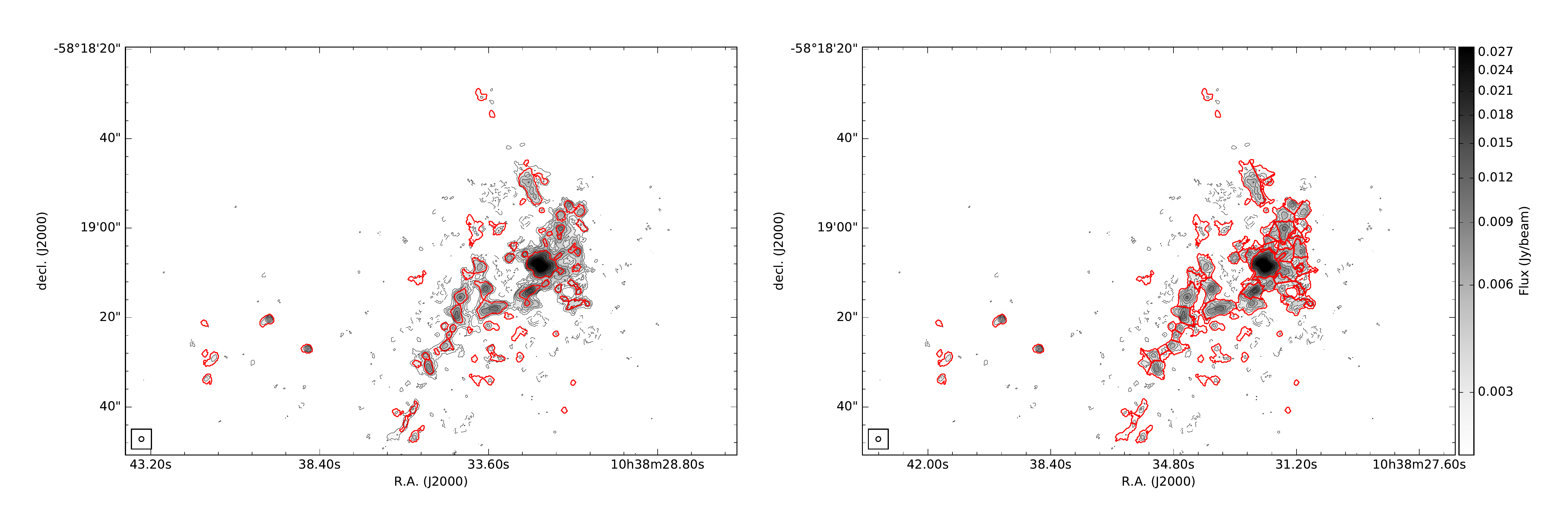}
\caption{
{\it (a) Left:} Cores found with the dendrogram method using our
fiducial criteria: $F_{\rm min}=4\sigma$, $\delta=1\sigma$ and $S_{\rm
  min}=0.5 S_{\rm beam}$. The image is shown in gray scale overlaid on
black contours starting from 4$\sigma$ and increasing in steps of
2$\sigma$. The red contours indicate the boundaries of the detected
cores. {\it (b) Right:} Same as (a), but now showing the results of
the clumpfind method. The criteria are $F_{\rm min}=4\sigma$,
$\Delta=1\sigma$ and $S_{\rm min}=0.5 S_{\rm beam}$.}\label{fig:2}
\end{figure*}

\subsection{Core Mass Estimation}

We estimate core masses by assuming optically thin thermal emission
from dust. The total mass surface density corresponding to a given specific
intensity of mm continuum emission is
\begin{eqnarray}
\Sigma_{\rm mm} & = & 0.369 \frac{F_\nu}{\rm mJy}\frac{(1\arcsec)^2}{\Omega} \frac{\lambda_{1.3}^3}{\kappa_{\nu,0.00638}}
 \nonumber\\
 & & \times  \left[{\rm exp}\left(0.553 T_{d,20}^{-1}
  \lambda_{1.3}^{-1}\right)-1\right]\:{\rm g\:cm^{-2}}\\
 &\rightarrow & 0.272 \frac{F_\nu}{\rm mJy}\frac{(1\arcsec)^2}{\Omega}\:{\rm g\:cm^{-2}}\nonumber,
\label{eq:Sigmamm}
\end{eqnarray}
where $F_{\nu}$ is the total integrated flux over solid angle
$\Omega$, $\kappa_{\nu,0.00638}\equiv\kappa_\nu/({\rm
  6.38\times10^{-3}\:cm^2\:g}^{-1})$ is the dust absorption
coefficient, $\lambda_{1.3}=\lambda/1.30\:{\rm mm}$ and
$T_{d,20}=T_d/20\:{\rm K}$ with $T_d$ being the dust temperature. To
obtain the above fiducial normalization of $\kappa_\nu$, we assumed an
opacity per unit dust mass $\kappa_{\rm 1.3mm,d}=0.899\: {\rm cm^2
  g}^{-1}$ (moderately coagulated thin ice mantle model of Ossenkopf
\& Henning 1994), which then gives $\kappa_{\rm 1.3mm}= {\rm
  6.38\times10^{-3}\:cm^2\:g}^{-1}$ using a
gas-to-refractory-component-dust ratio of 141 (Draine 2011). The
numerical factor following the $\rightarrow$ in the final line shows
the fiducial case where $\lambda_{1.3}=1$ and $T_{d,20}=1$.

Note that since we do not have detailed temperature information for
each source, for simplicity we have adopted an uniform value of
$T_{d}=20\:$~K for all cores in our fiducial analysis. Such
temperatures are expected to be representative of average temperatures
in protostellar cores (e.g., Zhang \& Tan 2015). However, we recognize
that somewhat warmer temperatures may result either from strong
external heating by nearby, luminous sources in the embedded
protocluster or by stronger than average internal heating in
protostellar cores. On the other hand, the temperature could be lower
in prestellar or early-stage protostellar cores. If temperatures of
15~K or 30~K were to be adopted, then the mass estimates would differ
by factors of 1.48 and 0.604, respectively.


Given the above values of $\Sigma_{\rm mm}$, then the core mass is 
\begin{eqnarray}  
M & = & \Sigma_{\rm mm} A = 0.113 \frac{\Sigma_{\rm mm}}{\rm g\:cm^{-2}} \frac{\Omega}{(1\arcsec)^2} \left(\frac{d}{\rm 1\:kpc}\right)^2  \:M_\odot\\
 & \rightarrow & 0.192 \frac{F_\nu}{\rm mJy} \left(\frac{d}{\rm 2.5\:kpc}\right)^2  \:M_\odot\nonumber
\end{eqnarray}
where $A$ is the projected area of the core, $d$ is the source
distance, and the final evaluation is for fiducial temperature
assumptions of 20~K (following eq.~\ref{eq:Sigmamm}). Thus the
$1\sigma$ noise level in the image corresponds to a core mass of
$\sim0.1\:M_\odot$.



Overall, we estimate absolute mass uncertainties of about a factor of
two, which we expect to be caused mostly by temperature
variations. Relative core mass estimates will be somewhat more
accurate, although still potentially with uncertainties of this
magnitude due to core to core temperature and opacity variations.

\subsection{Core Flux Recovery and Completeness Corrections}\label{S:recovery}

We calculate two corrections factors that are needed to estimate a
``true'' CMF from a ``raw'' observed CMF. First, since both dendrogram
and clumpfind methods adopt a threshold value (i.e., 4$\sigma$) and
pixels below this level are not assigned to any core structures, we
expect the estimated core flux (i.e., mass) is a fraction of the true
flux. We estimate the flux recovery fraction, $f_{\rm flux}$, as a
function of true core mass by carrying out experiments of artificial
core insertion into the {\it ALMA} images. These same experiments also
allow us to assess the second factor, i.e., the number recovery
fraction, $f_{\rm num}$, again as a function of true input core mass.
These correction factors are also expected to depend on core density
profile and the local clump environment, e.g., degree of crowding.


We adopt the following methods for these experiments of artificial
core insertion and recovery. The artificial cores are assumed to have
the same shape as the synthesized beam, i.e., the limiting case
appropriate for small, unresolved cores. The locations of the
artificial cores are chosen randomly, but with a probability density
that is scaled to match the flux profile we derive from the 7-m array
image, which has the effect of placing more cores in crowded
regions. In each experiment, we insert 10 cores (i.e., $\sim$10\% of
the total number to avoid excessive blending) of a given total flux,
i.e., of a given mass. We run the core detection algorithms to
determine the average flux levels recovered in detected cores and the
probability for artificial cores of a given mass to be found.  This is
repeated 30 times to obtain a large sample for more accurate
estimates.

With $f_{\rm flux}(M)$ estimated in this way, we then first transform
the raw CMF into a flux-corrected CMF, which involves estimating the
average (median) flux correction factor for a given observed
mass. Then, given our estimate of $f_{\rm num}(M)$, we transform the
flux-corrected CMF into an estimate of the true CMF, i.e., by assuming
the completeness correction factor at a given mass is equal to the
inverse of $f_{\rm num}$. The derived forms of $f_{\rm flux}(M)$ and
$f_{\rm num}(M)$ are shown in the next section for our fiducial case.



\section{Results}\label{S:results}

\subsection{1.3~mm Continuum Image}

Figure~\ref{fig:1} presents the $1.3\:$mm continuum map constructed
with the 7-m array data in the top left panel, 12-m and 7-m array combined
data in the top right panel, and the highest resolution combined image in
the bottom panel. The image with only 7-m data reveals two main
filaments: a northern one with a NE--SW orientation and a southern one
with a NW--SE orientation. These two filaments converge at a clump
with bright mm continuum emission. Several other isolated clumps are
also revealed. The southern filament and central hub are further
resolved into a cluster of dense cores. The image combining all data
has a spatial dynamic range that recovers structures from
$\sim1\arcsec$ to $\sim20\arcsec$.

Figure~\ref{fig:2} shows the high resolution ($\sim1\arcsec$) 1.3~mm
continuum image with the core boundaries overlaid for both the
dendrogram and clumpfind methods. Inspection of these images allows
one to assess how the core identification algorithms operate on the
imaging data. One sees cores with a range of sizes, some being many
times the size of the beam. Note that the central, brightest and most
massive ``core'' is identified in a similar way with both
algorithms. However, we expect that there is a high probability that
such massive, large area ``cores'' will appear fragmented when imaged
at higher angular resolution (see also \S\ref{S:discussion}).

Another feature revealed by Figure~\ref{fig:2} is clumpfind's method
of partitioning all the flux above the minimum threshold contour
level. This is to be contrasted with the method adopted by the
dendrogram algorithm, with the effect being to tend to make the cores
identified by clumpfind more massive than their dendrogram
counterparts.

\subsection{The Core Population and CMF}

\begin{figure}[ht!]
\epsscale{1.2}\plotone{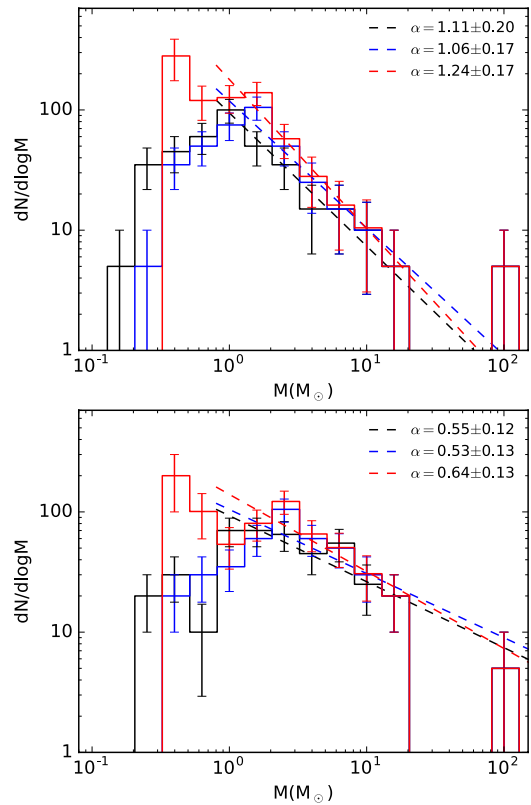}
\caption{
{\it (a) Top:}
CMF for the dendrogram method. The original CMF is shown in black and
after mass (flux) correction for each core is shown in blue. The blue
CMF is then corrected for the number recovery fraction, as illustrated
in red. 
The dashed lines in black, blue and red show the best power law fit
result for the high-mass end ($M>0.8\:M_{\odot}$) for the
corresponding CMFs. 
%
{\it (b) Bottom:} As (a), but now for the clumpfind method.
}\label{fig:3}
\end{figure}

\begin{figure}[ht!]
\epsscale{1.2}\plotone{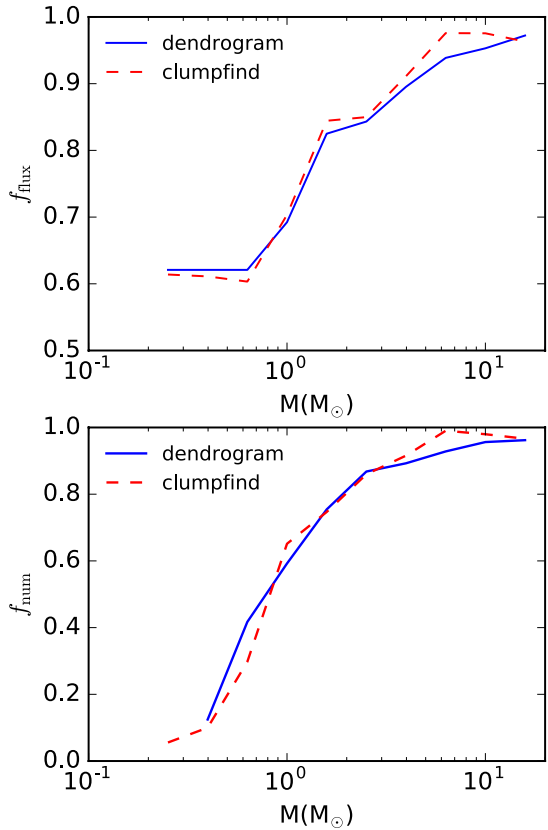}
\caption{
{\it (a) Top:} Flux recovery fraction, $f_{\rm flux}$, versus core
mass, $M$, for dendrogram and clumpfind algorithms, as labelled.
{\it (b) Bottom:} Number recovery fraction, $f_{\rm num}$, versus core
mass, $M$, for dendrogram and clumpfind algorithms, as
labelled.}\label{fig:corr}
\end{figure}

In Figure~\ref{fig:3} we show the ``raw'' CMFs (black histograms)
derived from our fiducial dendrogram (top panel) and clumpfind (bottom
panel) methods. The fiducial dendrogram method ($F_{\rm min}=4\sigma$,
$\delta=1\sigma$, $S_{\rm min}=0.5S_{\rm beam}$) identifies 76 cores,
while the fiducial clumpfind method ($F_{\rm min}=4\sigma$,
$\Delta=3\sigma$, $S_{\rm min}=0.5 S_{\rm beam}$) finds 83
cores. 
Note, we adopt uniform binning in ${\rm log}\: M$, with 5 bins per
dex. Poisson counting errors are shown for each
bin. Figure~\ref{fig:3} also displays the flux corrected CMFs (blue
histograms, with errors again estimated as a Poisson value) and
subsequently number corrected, i.e., ``true,'' CMFs (red histograms,
with error assumed to be the same fractional value as in the blue
histograms), for each case. The fitting of power law functions to the
high-mass end of the CMFs is discussed below.

The correction factors used in Figure~\ref{fig:3} are shown in
Figure~\ref{fig:corr}. The flux correction factor, which is based on
median values of $f_{\rm flux}$ (excluding values $>1$, which we
attribute to false assignments; and extrapolating with constant
values for $M\lesssim 0.3\:M_\odot$), rises from about 0.6 at the
low-mass end (when cores are detected) to close to unity at the
high-mass end. The values of $f_{\rm flux}$ for dendrogram and
clumpfind are similar to each other, with clumpfind recovering
slightly more flux over most of the mass range.


The number recovery fractions, $f_{\rm num}$, show a larger dynamic
range, rising from $\sim0.1$ at the low-mass end to near unity at the
high-mass end (these remain slightly less than one due to the
possibility of blending with existing massive cores). Again, the
values of this correction factor are similar for both dendrogram and
clumpfind. We estimate that we are about 50\% complete by number for
$\sim1\:M_\odot$ cores. The direct effect of the number correction can
be seen by comparing the blue and red histograms in Fig.~\ref{fig:3}.



We characterize the high-end ($>0.8\:M_\odot$, i.e., starting with the
bin centered on $1\:M_\odot$) part of the raw dendrogram CMF by
fitting a power law of the form given by equation~(\ref{eq:pl}).  We
find $\alpha=1.11\pm0.20$. Fitting the same mass range for the flux
corrected CMF yields $\alpha=1.06\pm0.17$, while that for the fully
(flux and number) corrected, i.e., ``true'', CMF yields
$\alpha=1.24\pm0.17$. Thus these correction factors have only a modest
impact on the shape of the CMF for $M\gtrsim 0.8\:M_\odot$, with the
true CMF being slightly steeper than the raw CMF, mostly due the
effects of the number correction.

We note that there is sparse sampling of the high-mass end of the CMF,
i.e., there is a single, massive ($\sim 100\:M_\odot$) ``core.'' Our
fitting method, which we note minimizes $\chi^2$ in log space, treats
the empty bins as effective upper limits.
However, if we were to exclude this source and fit the CMF only over
the range from 0.8 to $\sim20\:M_\odot$, then we would derive
$\alpha=1.11\pm0.22$ and $\alpha=1.15\pm0.17$ for the raw and true
CMFs, respectively, i.e., there is only a very minor
effect.

Inspection of the true CMF indicates that the power law behavior may
continue down to lower masses. If we fit to the range
$M\gtrsim0.3\:M_\odot$, we derive a moderately shallower value of
$\alpha=0.83\pm0.11$. From these results, we see that there is
potential evidence for a break in the CMF near $1\:M_\odot$, but that
a single power law is still a reasonable description of the flux and
number corrected, i.e., true, CMF across most of the mass range
probed, i.e., from $\sim0.3\:M_\odot$ to $\sim100\:M_\odot$.


For the CMF resulting from the fiducial clumpfind algorithm, the power
law description of the raw CMF also appears potentially valid for
$M\gtrsim0.8\:M_\odot$. For this we derive $\alpha=0.55\pm0.12$, which
is significantly shallower than the $1.11\pm0.20$ derived over the
same mass range for the dendrogram raw CMF. Thus, note, there are a
larger number of massive cores found with the clumpfind method than
with the dendrogram method. Then, on applying the flux and number
corrections, the ``true'' CMF found via clumpfind displays a local
peak at about $2.5\:M_\odot$, but with numbers of lowest-mass cores
still potentially rising slowly. If we attempt the same uniform metric
of a single power law fit above $0.8\:M_\odot$, then we find
$\alpha=0.64\pm0.13$.
If we fit only from the bin containing the true CMF peak and extending
to higher masses, then we find $\alpha=0.78\pm0.14$, which is still
shallower than the equivalent dendrogram result.

Thus we see that whether or not there is a peak or break defining a
characteristic mass in the CMF depends on the method of core
identification used and whether or not completeness corrections are
applied. In particular, while the two methods find similar number of
cores, we can explain the differences in their final CMFs mostly as a
result of how mass is then assigned to the identified structures. As
discussed above, clumpfind partitions all the flux above a given
threshold to the sources, while dendrogram does not, i.e., its cores
sit on plateaux that are described by branches in its structural
decomposition.

\begin{figure*}[ht!]
\epsscale{1.18}\plotone{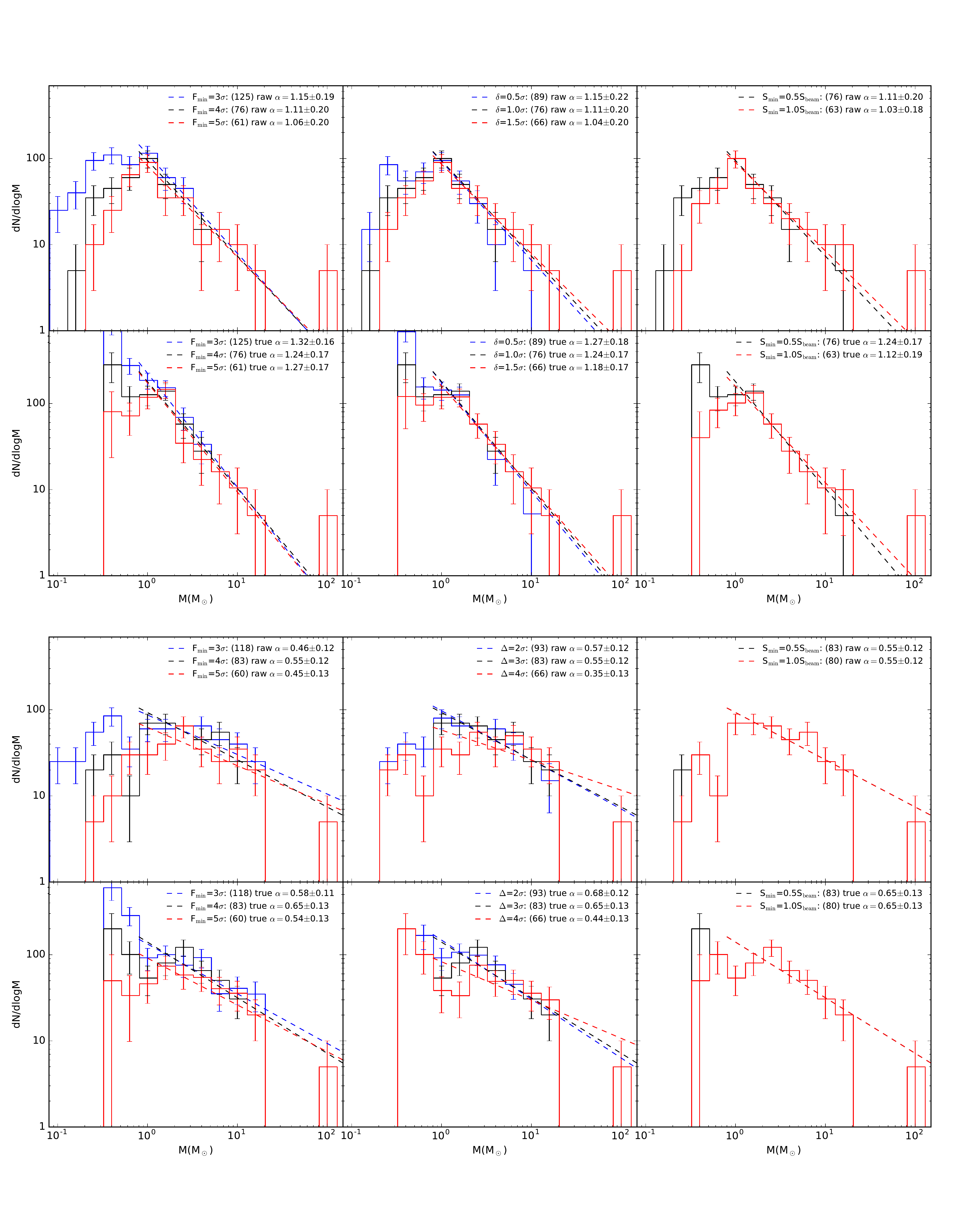}
\caption{
Raw CMFs derived with results from the dendrogram method shown on the
top panels, and the clumpfind method on the bottom panels. For each
algorithm we show different results by varying $F_{\rm min}$, $\delta$
(for dendrogram and $\Delta$ for clumpfind) and $S_{\rm min}$
(columns, left to right). In each panel, the results with different
parameter selections are illustrated in different colors (see
text). The number in the brackets denotes how many cores are
detected. Also shown is the power law index, $\alpha$, from fitting
the high-mass end ($M>0.8\:M_{\odot}$ for both dendrogram and
clumpfind).}\label{fig:all}
\end{figure*}

The values of high-end slopes of the CMFs are relatively unaffected by
the application of the completeness corrections.
We note that the stellar IMF at $\gtrsim 1\:M_\odot$ also follows a
power law form with $\alpha\simeq 1.35$ (Salpeter 1955), and this
value is very similar to those seen in the dendrogram CMFs, while the
clumpfind CMFs are shallower. As previous studies of more local
regions have found (see \S\ref{S:intro}), this may indicate that core
to star formation efficiency is relatively constant with increasing
mass, at least over the range of masses that is effectively probed
here, i.e., from $\sim1$ to $\sim100\:M_\odot$. The outflow and
radiative feedback models of Tanaka et al. (2017) for star formation
in clumps with $\Sigma_{\rm cl}\simeq 1\:{\rm g\:cm}^{-2}$, i.e., the
value most relevant to G286, predict that these efficiencies should
drop from $\epsilon=$0.48 to 0.37 as the stellar mass increases from
$5\:M_\odot$ to $40\:M_\odot$, i.e., as core masses increase from
about $10\:M_\odot$ to about $100\:M_\odot$. Such a relatively small
change in $\epsilon$ is still compatible with the results we have
presented, since they lack significant numbers of cores $>20\:M_\odot$
to place very stringent constraints in this regime. Other caveats
should also be considered that may affect the derived CMFs, including
possible systematic temperature variations with increasing continuum
flux, i.e., if brighter cores are warmer, we will have overestimated
their masses. However, with the data in hand, it is not currently
possible to assess how important this effect may be.

In Figure~\ref{fig:all} we show the dependence of the CMFs that result
from varying the three main parameters associated with each core
identification method. We focus on the total core numbers found, the
shape of the raw and true CMFs, and the high-end slope of the power
law fits. In relation to the fiducial dendrogram method, if we lower
the minimum threshold to $F_{\rm min}=3\sigma$, 125 cores are now
found (total core numbers are listed in parentheses in the legend in
Fig.~\ref{fig:all}), with the increase mostly being for sub-solar mass
cores. If we set $F_{\rm min}=5\sigma$, then only 61 cores are
recovered. Varying $\delta$ to $0.5\sigma$ or $1.5\sigma$ has a more
modest effect, as does increasing the minimum size of a core to 1 beam
area. We see from comparing the raw CMFs and their derived values of
$\alpha$ that the shape above $1\:M_\odot$ is relatively robust to
these variations. In fact, we note that all the variation we see in
$\alpha$ of these raw CMFs due to different dendrogram parameter
choices is smaller than the uncertainty arising from Poisson counting
statistics in this fiducial estimate. The completeness-corrected
``true'' CMFs found by the different dendrogram methods are generally
very similar to one another if one restricts attention to $M\gtrsim
1\:M_\odot$, where the power law fits are always found to be slightly
steeper than those of the raw CMFs. However, the shapes of these true
CMFs below $1\:M_\odot$ are quite strongly affected by the choice of
core definition within the dendrogram framework. This can affect
whether or not a characteristic core mass is seen in the CMFs.

We have seen that the fiducial clumpfind method yields similar core
numbers as the dendrogram analysis. Figure~\ref{fig:all} shows that
this is also true if we consider variations in its parameters $F_{\rm
  min}$ and $\Delta$, in correspondence with the variations of the
equivalent dendrogram parameters. However, unlike dendrogram,
clumpfind does not see a significant reduction in the numbers of cores
found if the minimum core size is doubled. Again, most values of the
high-end $\alpha$ of these raw and true CMFs are similar to the
fiducial values of their respective cases, i.e., 0.55 and 0.65, with
only the $\Delta=4\sigma$ case yielding significantly shallower
slopes.


\begin{figure*}[t]
\epsscale{1.0}\plotone{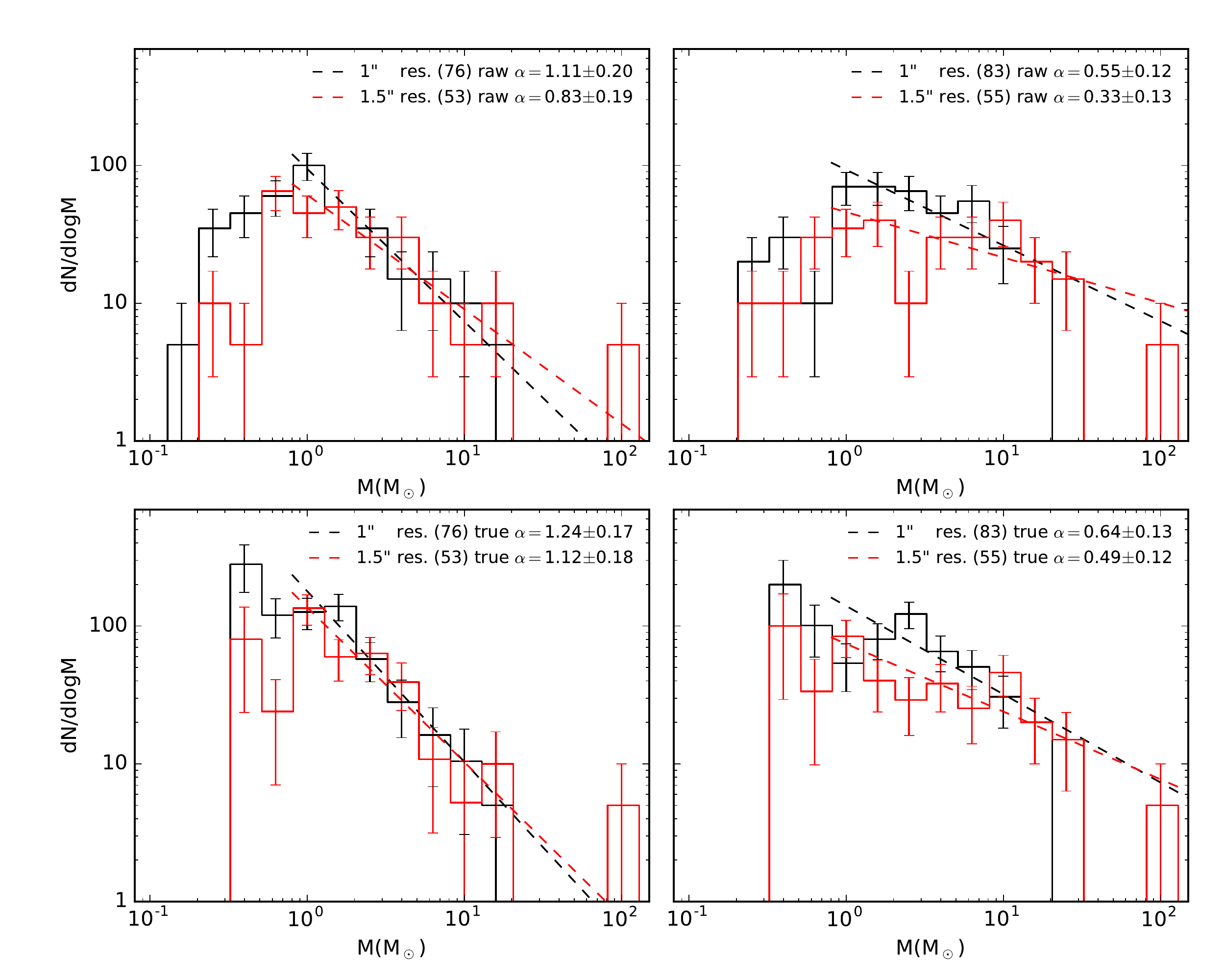}
\caption{
CMFs derived for images with lower spatial resolution, i.e.,
``1.5\arcsec'' (actually 1.62\arcsec$\times$1.41\arcsec), shown as red
histograms and fitted power laws. These are compared to the fiducial
results from analysis of the ``1\arcsec'' images (actually
$1.07\arcsec\times1.02\arcsec$), shown in black. {\it Top left:} Raw
CMFs with the fiducial dendrogram method. {\it Bottom left:}
Completeness-corrected true CMFs with the fiducial dendrogram
method. {\it Top right:} Raw CMFs with the fiducial clumpfind method.
{\it Bottom right:} Completeness-corrected true CMFs with the fiducial
clumpfind method.}\label{fig:res}
\end{figure*}

Next, we examine how the CMFs vary if the analyzed image has a
lower angular resolution of $\simeq1.5\arcsec$. Figure~\ref{fig:res}
compares the raw CMFs derived from the $1\arcsec$ and
$\simeq1.5\arcsec$ images. As expected, core masses tend to shift to
higher values when identified from the lower resolution image. This
leads to a flattening in the shape of the high-end CMFs, i.e., a
reduction in the derived values of $\alpha$, which can be quite
significant, i.e., $\Delta\alpha\simeq - 0.3$ for the raw CMF found by
the fiducial dendrogram method. However, after completeness
corrections are applied, the effect on $\alpha$ is more modest.
These results indicate that even the high-end part of the CMFs can
vary somewhat as the resolution is changed, and the trend may continue
in the opposite direction if one were to image at higher
resolutions. Indeed, this is expected if the more massive, larger
cores are seen to fragment at significant levels when imaged at higher
resolution. Such cores are known to fragment to some extent, although
there are observed cases of quite limited fragmentation (e.g.,
Csengeri et al. 2017). This effect should be kept in mind when
comparing CMFs derived from protoclusters that are observed with
different resolutions, e.g., as may occur due to being at different
distances.

\begin{figure*}[t]
\epsscale{1.0}\plotone{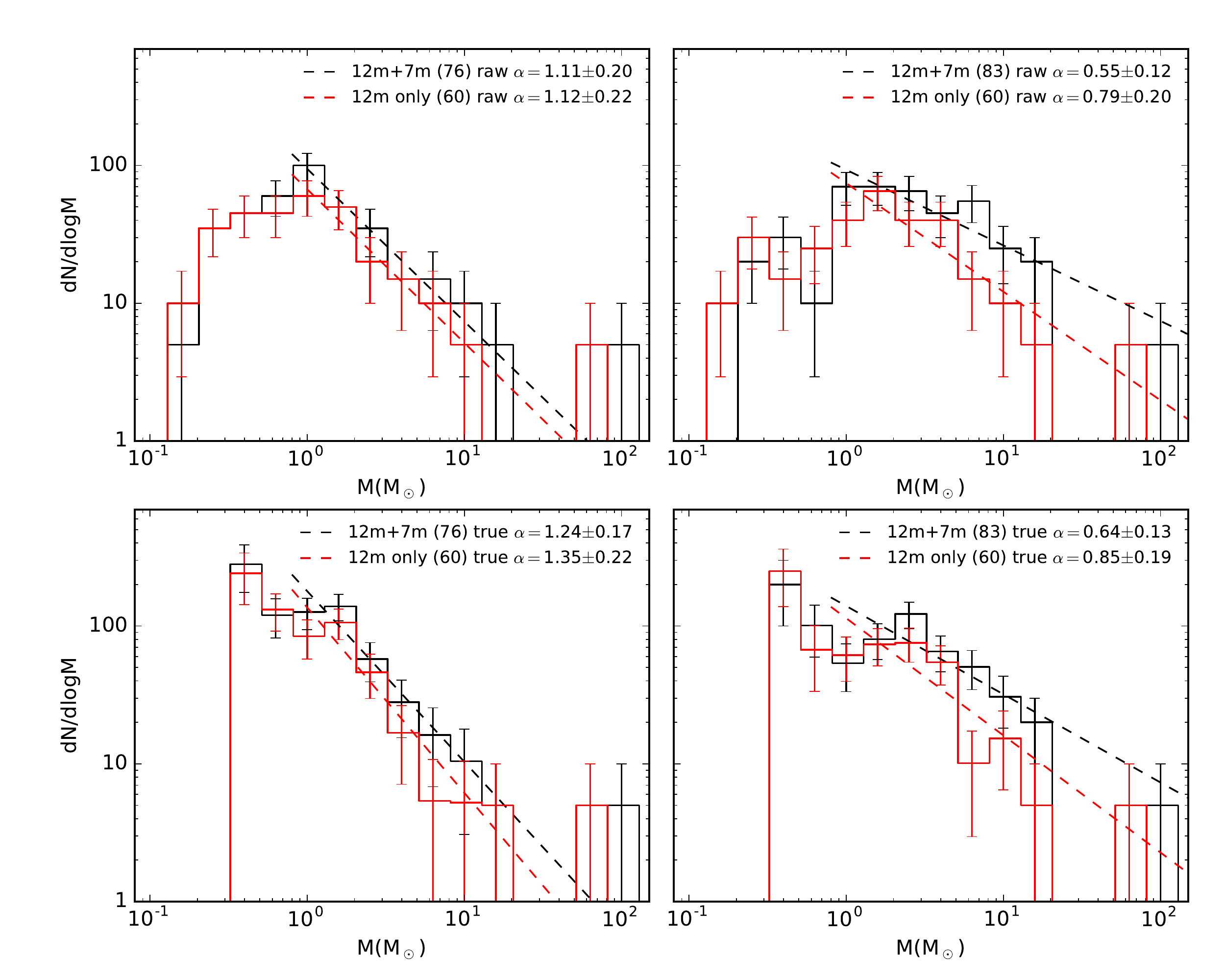}
\caption{
CMFs derived for images derived from only the 12m-array data, shown as
red histograms and fitted power laws. These are compared to the
fiducial results from analysis of our 12m + 7m array combined images,
shown in black. {\it Top left:} Raw CMFs with the fiducial dendrogram
method. {\it Bottom left:} Completeness-corrected true CMFs with the
fiducial dendrogram method. {\it Top right:} Raw CMFs with the
fiducial clumpfind method.  {\it Bottom right:} Completeness-corrected
true CMFs with the fiducial clumpfind method.}\label{fig:12m}
\end{figure*}

Finally, in Figure~\ref{fig:12m} we examine how the CMFs vary if the
analyzed image is lacking the larger spatial scales obtained from the
7m-array data. Such an analysis is useful for understanding how the
results of other observational programs that measure CMFs without such
data may be affected. Our 12m only image has an rms noise level of
0.47~mJy~beam$^{-1}$. For the dendrogram method we find that the CMF
derived from the 12m only image contains slightly fewer cores (60)
than found in the combined image (76), but has a high-end power law
slope index that is very similar. For the completeness-corrected CMF
the 12m-array only CMF has a high-end power law index that is about
0.1 steeper than that derived from the 12m + 7m image. Similar results
are also found for clumpfind derived raw and true CMFs, with the
difference now being about 0.2 in the magnitude of $\alpha$. Thus the
value of the high-end power law slope of the true CMF appears to be
slightly over estimated if the image is lacking the larger spatial
scales provided by 7m-array data.

\section{Discussion and Conclusions}\label{S:discussion}

We have studied the CMF in the central region of the massive
protocluster G286.21+0.17, with cores identified by their 1.3~mm dust
continuum emission in a high spatial dynamic range image observed with
the 7-m and 12-m arrays of ALMA. We explored the effects of using two
different core identification algorithms, dendrogram and clumpfind,
including a systematic study of the effects of varying their three
main core selection parameters. We also examined the effects of
varying angular resolution and largest recovered angular scale of the
analyzed continuum image.

Our fiducial methods, including flux and number corrections estimated
by artificial core insertion and recovery, yield CMFs that show
high-end ($M\gtrsim 1\:M_\odot$) power law indices of
$\alpha=1.24\pm0.17$ for dendrogram and $0.64\pm0.13$ for
clumpfind. These results are quite robust to variations of choices of
core selection parameters. 

With the dendrogram method, which we consider to be preferable to
clumpfind as a means for identifying and characterizing cores that are
embedded in a clump environment, these power law indices are similar
to the Salpeter stellar IMF index of 1.35. This further strengthens
the case of a correspondence between CMF and IMF seen in local
regions, but now in a more distant, massive protocluster. As discussed
in \S\ref{S:intro}, such a correspondence is a general feature and/or
expectation of Core Accretion models of star formation, in contrast to
Competitive Accretion models. However, caveats remain, including
potential systematic changes in core temperature for brighter cores
and the fact that the measured CMF is expected to be composed of a
mixture of prestellar and protostellar cores, i.e., tracing different
evolutionary stages (see also discussion of Clark et al. 2007).


We do find that whether or not a peak is seen in the CMF near
$1\:M_\odot$ depends on which core finding algorithm is used, i.e.,
dendrogram or clumpfind, the choices of parameters associated with the
algorithm, and whether or not completeness corrections are carried
out. Thus we cannot make firm conclusions about the presence of a peak
or characteristic core mass near $1\:M_\odot$. Such a peak might be
expected if there is close correspondence of CMF shape with stellar
IMF shape. Our fiducial dendrogram result (see Fig.~\ref{fig:3}a)
shows only a very tentative hint of there being a break in the power
law description of the CMF to shallower slopes for masses $\lesssim
1\:M_\odot$.



We re-emphasize that the relation of the CMF identified purely from
sub-mm/mm dust continuum emission to the stellar IMF is uncertain. We
expect that many of the cores identified by these methods, being the
brighter cores, will be protostellar sources. Examples of massive
prestellar cores identified by their high levels of deuteration, i.e.,
via $\rm N_2D^+$ line emission, can show relatively weak mm continuum
emission, perhaps indicating that they are significantly colder than
their surrounding clump material (Kong et al. 2017a,b). For
constraining theoretical models, it is desirable to have a measure of
the PSCMF, and it remains to be seen how effective interferometric
studies of mm continuum emission in distant massive protoclusters are
at measuring this PSCMF (see, e.g., Fontani et al. 2009).

The observations carried out here also included $\rm N_2D^{+}$(3-2)
and $^{12}$CO(2-1), amongst other species. In a future paper, these
line data
will be analyzed to place better constraints on the PSCMF and its
relation to the CMFs presented here. We note that core finding methods
that also utilize molecular line emission may also make it easier to
break-up spatially confused structures.

Another caveat in the accuracy of CMF determination relates to the
effects of spatial resolution and the possibility of fragmentation of
identified ``cores'' into smaller structures as the resolution is
increased. Such an effect has been seen before in numerous studies,
but at varying levels (e.g., Beuther \& Schilke 2004; Bontemps et
al. 2010; Zhang et al. 2015; Csengeri et al. 2017). Cases of limited
fragmentation may indicate an important role for magnetic fields in
stabilizing the more massive cores (see, e.g., Kunz \& Mouschovias
2009; Tan et al. 2013; Fontani et al. 2016). Our investigation of how
the true dendrogram CMF varies as the resolution is changed from about
1.5\arcsec\ to 1\arcsec\ shows that there is a slight steepening of
the power law index, by about 0.1, as one goes to the higher
resolution. However, the size of this change is smaller than the
uncertainties arising solely from counting statistics, so larger
samples of cores are needed to verify this trend. Higher sensitivity
and higher angular resolution studies of the G286.21+0.17 are also
desirable to investigate the particular fragmentation properties of
the identified cores.

Taking the above caveats of CMF definition in mind, we still regard
characterization of the mm continuum image via identification of
discrete cores by specified, well-defined algorithms as a useful
exercise for assessing the fragmentation in the cloud and as a first
step for measuring the true CMF and, eventually, the
PSCMF. Furthermore, the same core finding algorithms can also be
applied to simulated molecular clouds to make a direct, statistical
comparison of their structures with those of real systems, and in this
way constrain the physics of star and star cluster formation.

\acknowledgements YC acknowledges a Graduate School Fellowship from
the University of Florida. JCT acknowledges NSF grants AST1312597 and
AST1411527. This paper makes use of the following ALMA data:
ADS/JAO.ALMA\#2015.1.00357.S. ALMA is a partnership of ESO
(representing its member states), NSF (USA) and NINS (Japan), together
with NRC (Canada), NSC and ASIAA (Taiwan), and KASI (Republic of
Korea), in cooperation with the Republic of Chile. The Joint ALMA
Observatory is operated by ESO, AUI/NRAO, and NAOJ. The National Radio
Astronomy Observatory is a facility of the National Science Foundation
operated under cooperative agreement by Associated Universities, Inc.

\end{document}